\documentclass[pre,twocolumn,showpacs,floatfix,amsmath,amssymb]{revtex4-1}

\usepackage{graphicx, color}
\usepackage{amsmath,amssymb}
\usepackage{verbatim}

\newcommand{\figOne}{
\begin{figure}[t]
	\centering
	\includegraphics[width=2.25in]{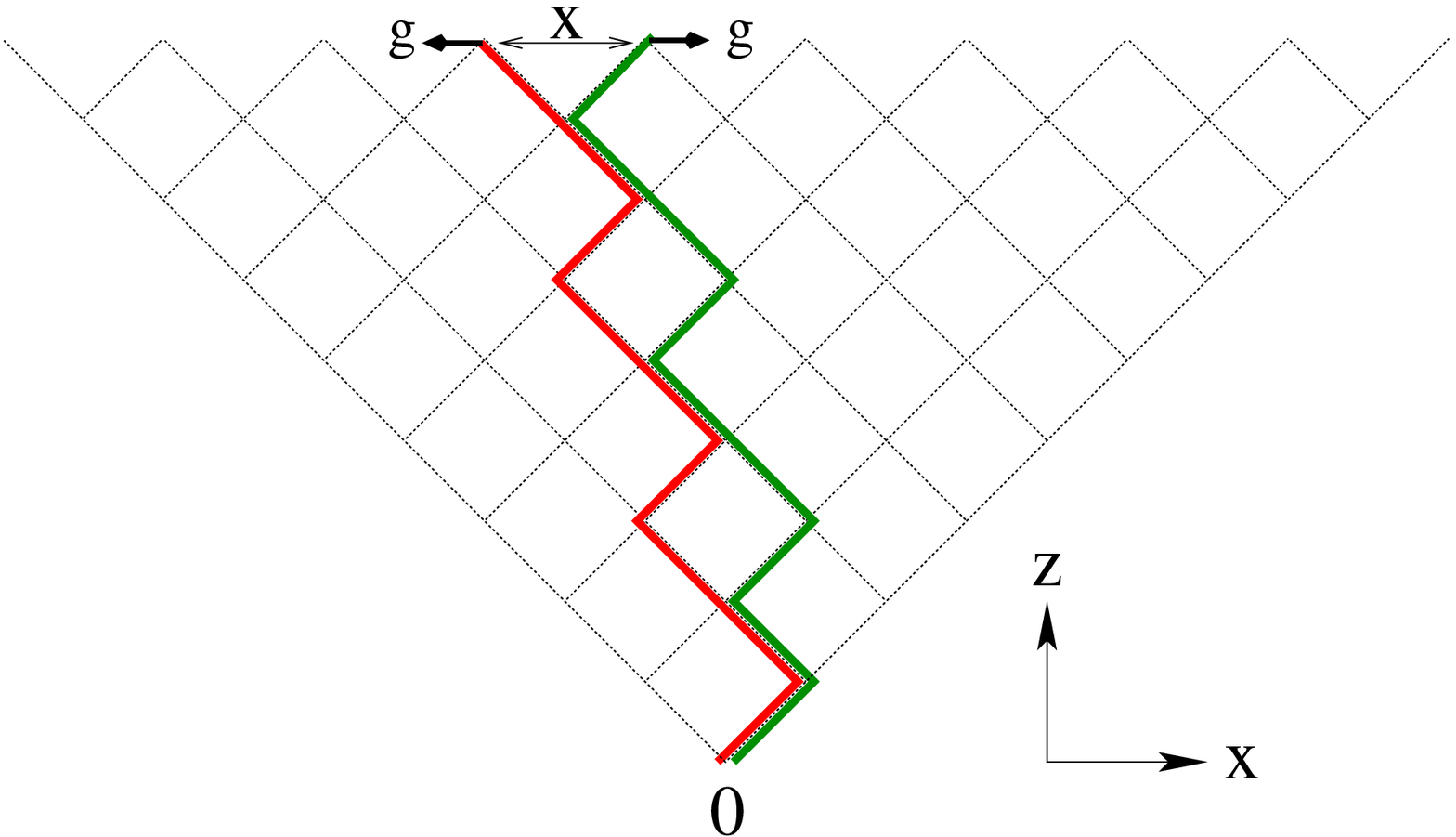}
	
	\caption{\label{fig:model} 
	(Color online) Schematic diagram of DNA unzipping. The strands of the DNA are shown by thick
	solid lines with a pulling force $g$ act along $x$ direction at end monomers. }

\end{figure}
}

\newcommand{\figTwo}{
\begin{figure*}[t]
	\centering
	\includegraphics[height=2.35in]{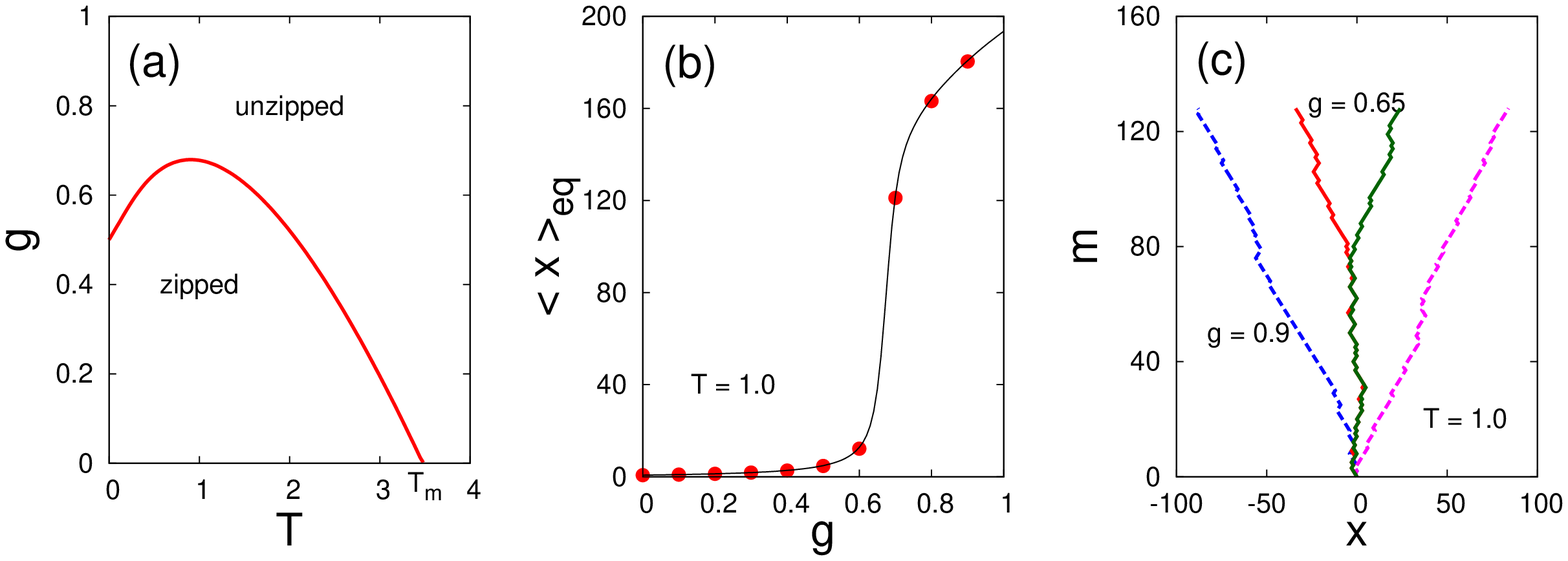}

	\caption{\label{fig:pd} (Color online) (a) Critical unzipping force as a function of
		temperature [Eq.~(\ref{eq:gc})]. (b) Force $g$ vs equilibrium average separation $\langle x
	\rangle_{eq}$ between the end monomers of the dsDNA at $T=1.0$. The line is from the exact
	transfer matrix approach and points are from the Monte Carlo simulations. (c) Typical
	equilibrium configurations of the dsDNA of length $N=128$ for force values $g=0.65$ (lies
	just below the phase boundary) and $g=0.9$ (far above the phase boundary) at $T=1.0$ obtained
	by using Monte Carlo simulations. The lines are the bonds between monomers of the strands.}

\end{figure*}
}

\newcommand{\figThree}{
\begin{figure}[b]
	\centering
	\includegraphics[width=3.0in]{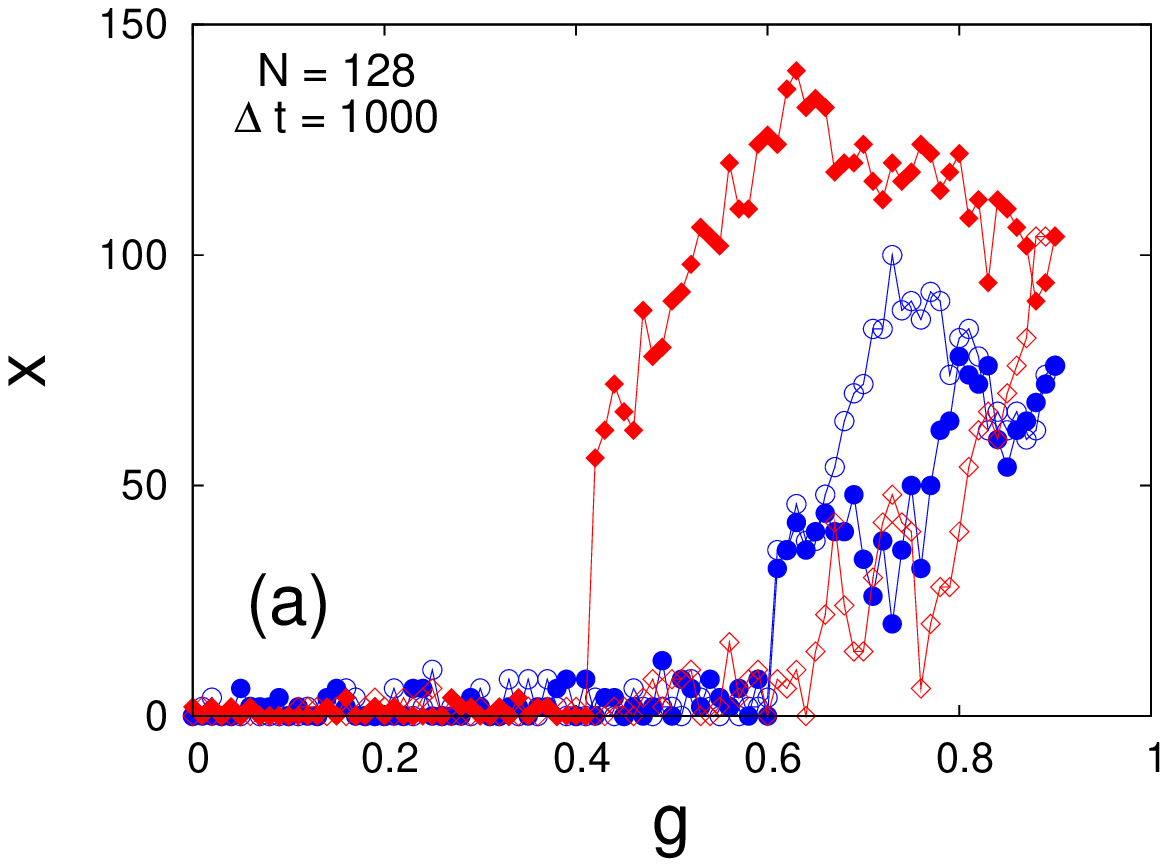}
	\includegraphics[width=3.0in]{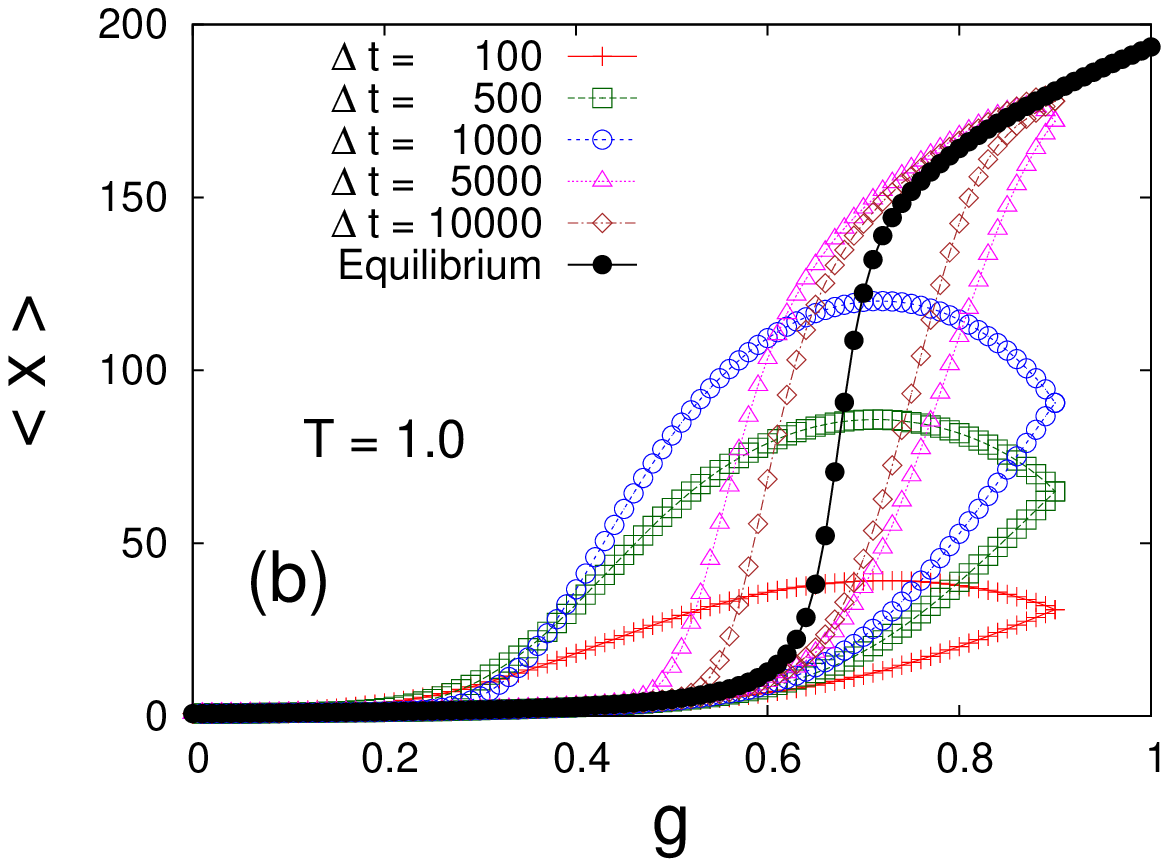}
	
	\caption{\label{fig:sam} (Color online) (a) The separation $x$ between the end monomers of
	the DNA as a function of force $g$ for the forward (unzipping) and the backward (rezipping)
	paths for two different realizations of a dsDNA of length $N=128$ at $T=1$ for the pulling
	rate $\Delta t = 1000$. The filled and the open symbols represent the forward and the
	backward paths, respectively. The work done over a complete cycle is negative (positive) for
	the realization shown by circles (diamond). (b) Hysteresis curves for different pulling
	rates. The equilibrium curve, shown by the filled circles, does not show any hysteresis. The
	paths are averaged over $10^5$ different realizations.  }

\end{figure}
}

\newcommand{\figFour}{
\begin{figure}[b]
	\centering
	\includegraphics[width=3.0in]{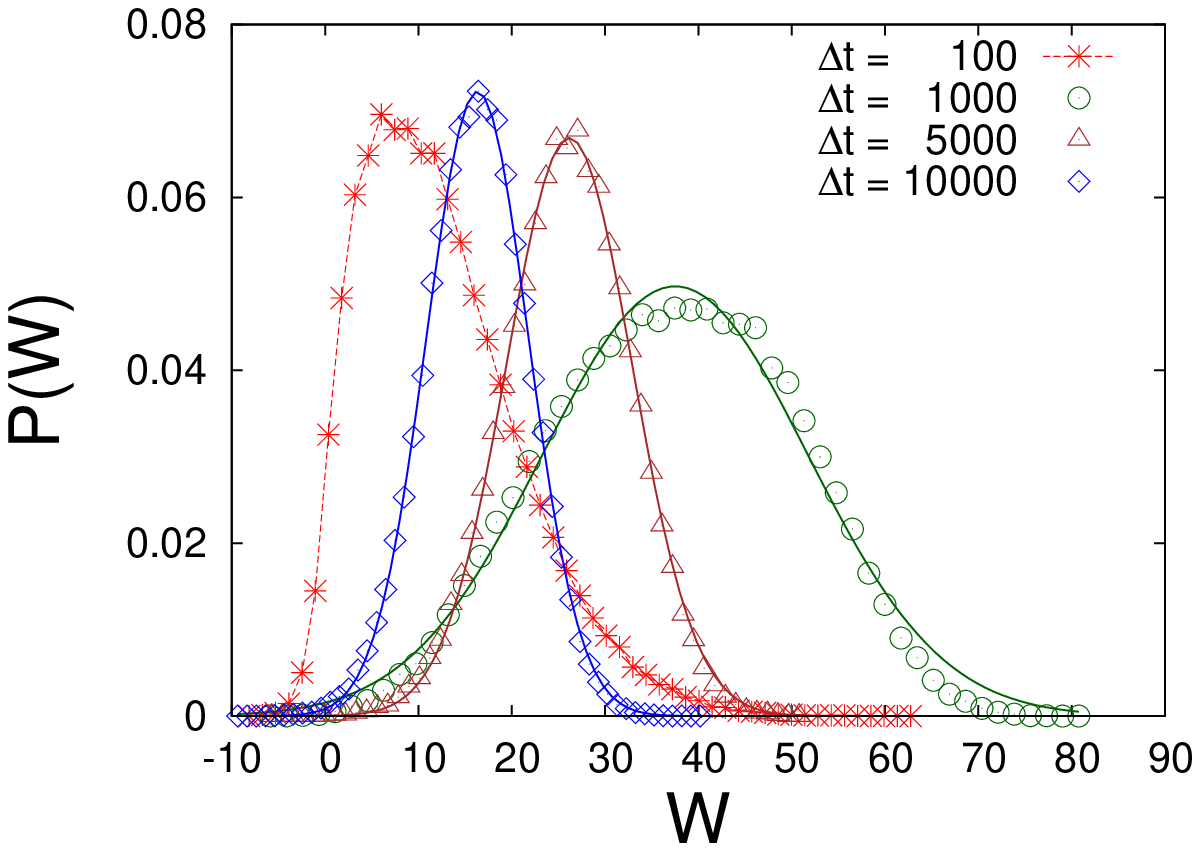}
	
	\caption{\label{fig:prob} (Color online) Probability distribution P(W), of the work performed
	during a complete 	unzipping and rezipping cycle for various $\Delta t$ values. The solid
	lines are the Gaussian fit to the distributions while the dashed line for $\Delta t = 100$ is
	guide to eyes.}

\end{figure}
}

\newcommand{\figFive}{
\begin{figure*}[t]
	\centering
	\includegraphics[height=5.0in]{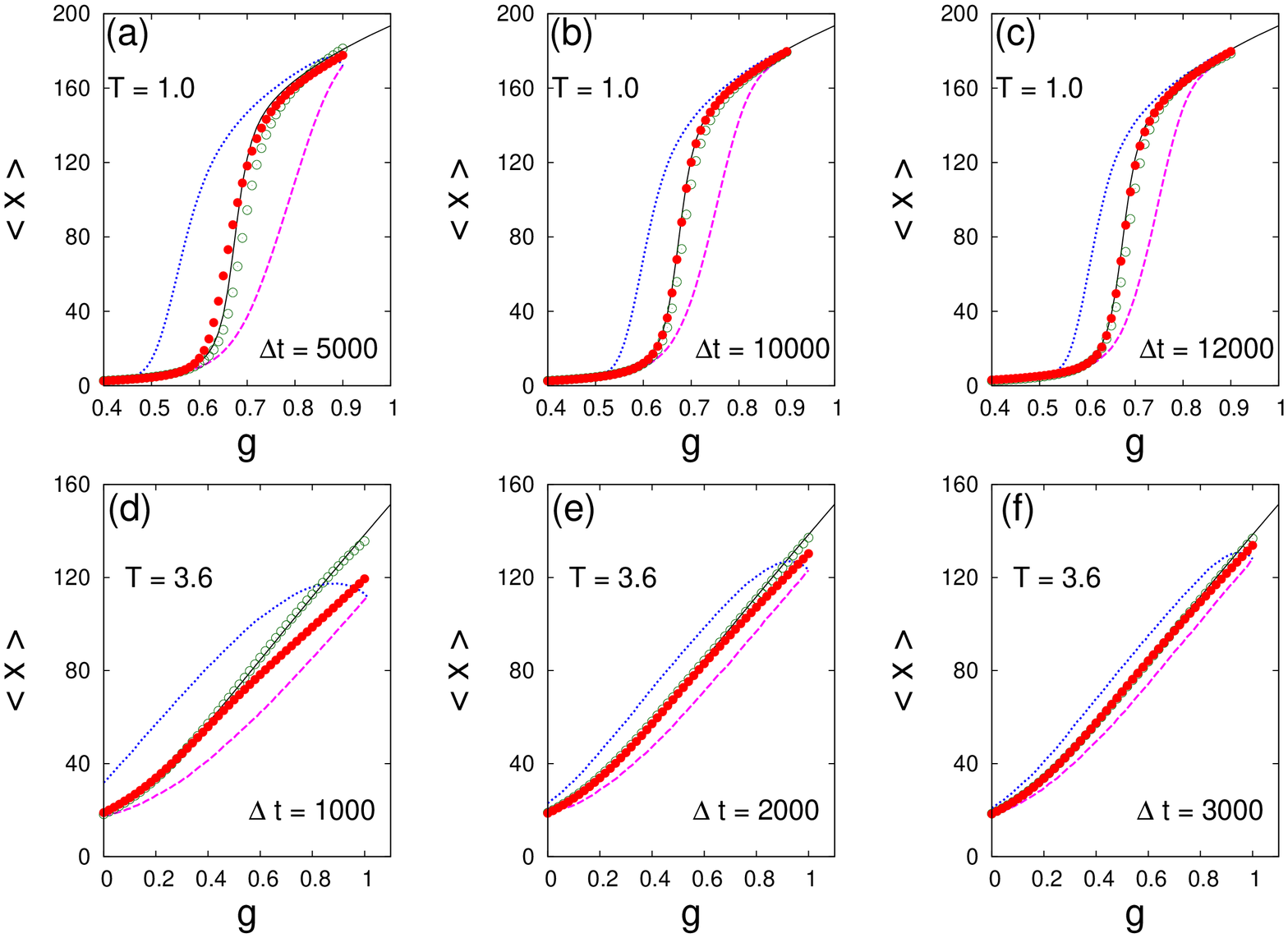}
	
	\caption{\label{fig:hyst} (Color online) Hysteresis obtained in unzipping and rezipping of a
	dsDNA for various pulling rates at two different temperatures, (a) to (c) at $T = 1.0 < T_m$
	and (d) to (f) at $T = 3.6 > T_m$. The forward (shown by dashed line) and the backward (shown
	by dotted line) paths are averaged over $10^5$ realizations for (a) and (b) while for cases
	(c) to (f) paths are averaged over $10^4$ realizations. The unfilled and filled circles
	represent the equilibrium curves obtained by using work theorem on nonequilibrium forward and
	backward paths, respectively. The solid lines represent the equilibrium curves obtained by
	using the exact transfer matrix technique. }

\end{figure*}
}

\newcommand{\figSix}{
\begin{figure*}[t]
	\centering
	\includegraphics[width=5.0in]{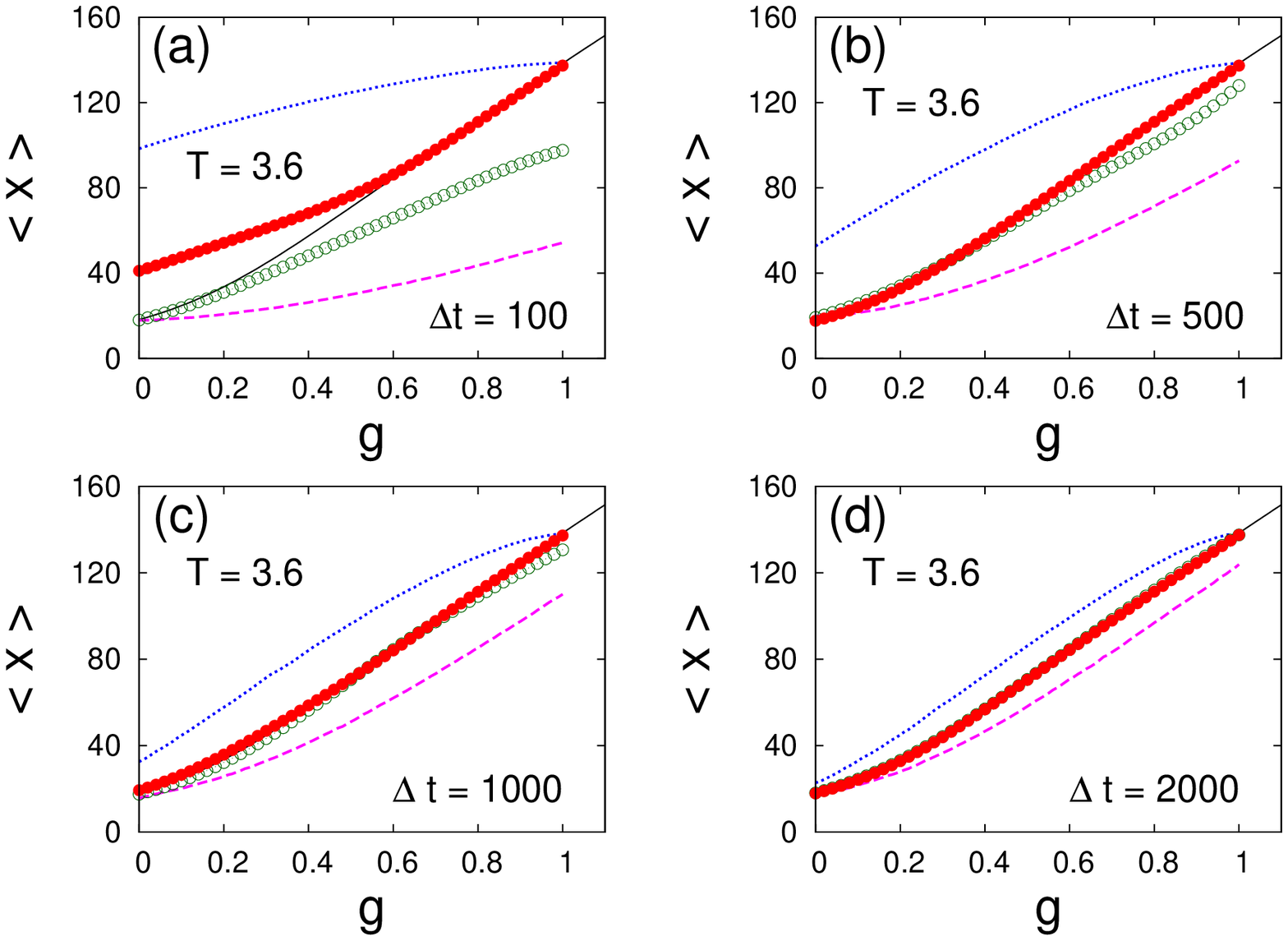}
	
	\caption{\label{fig:hysteq} (Color online) The equilibrium curves, shown by unfilled (filled)
	circles for the forward (backward) paths, obtained by using work theorem on nonequilibrium
	paths when the backward paths are also initially equilibrated at $g_m$.  The dashed and
	dotted lines are, respectively, the forward and backward nonequilibrium paths averaged over
	$10^4$ realizations at $T=3.6$ for various pulling rates. The solid line represent the
	equilibrium curve obtained by using the exact transfer matrix technique. }

\end{figure*}
}

\newcommand{\figSeven}{
\begin{figure}[t]
	\centering
	\includegraphics[width=3.0in]{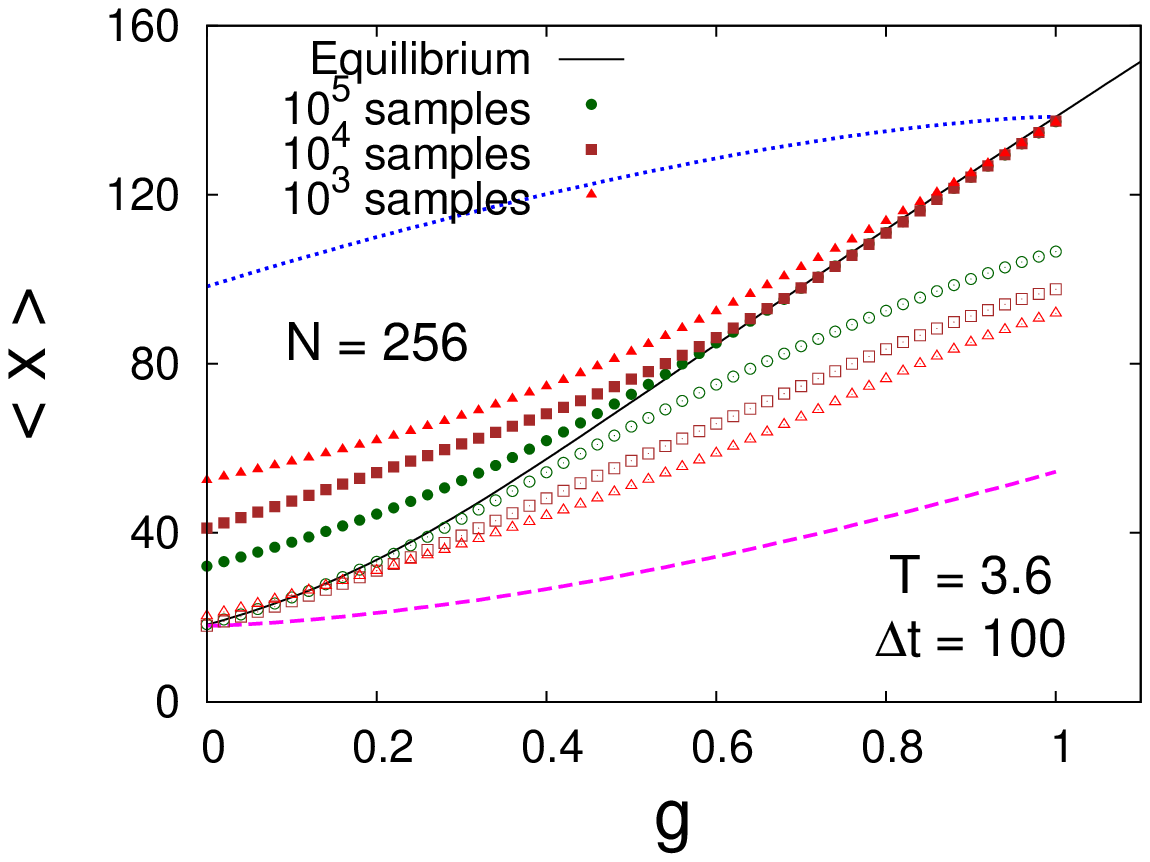}
	
	\caption{\label{fig:hysteq100} (Color online) Equilibrium $g$ versus $\langle x \rangle$
	isotherms obtained by using work theorem on $10^3$, $10^4$, and $10^5$ nonequilibrium
	trajactories for the pulling rate $\Delta t = 100$ at $T=3.6$. The unfilled (filled) circles
	represent the forward (backward) path. The solid lines represent the exact equilibrium curve.
	}

\end{figure}
}

\newcommand{\figEight}{
\begin{figure}[t]
	\centering
	\includegraphics[width=3.0in]{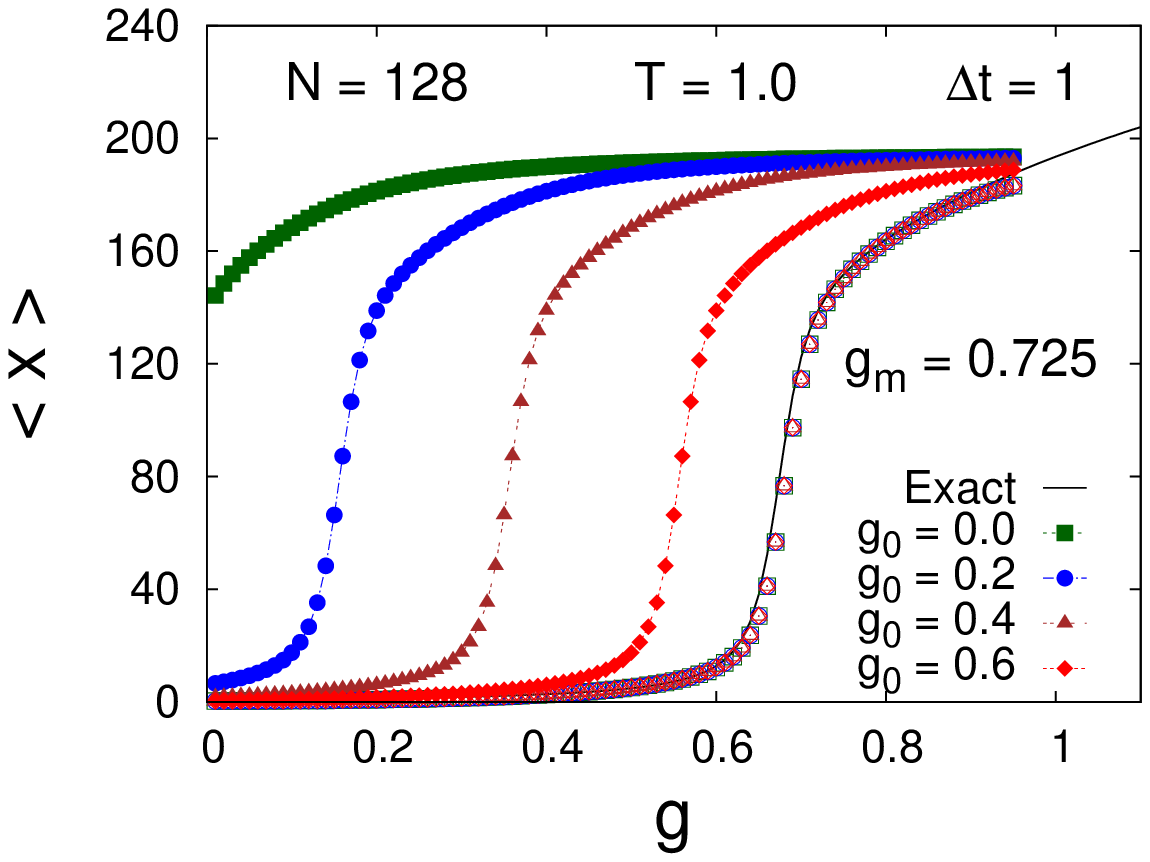}
	
	\caption{\label{fig:spont} (Color online) Equilibrium and nonequilibrium $g$ versus $\langle
	x \rangle$ isotherms obtained by using a work theorem on $10^4$ trajactories when the pulling
	force on the DNA decreases spontaneously from $g_m=0.725$ to various $g_0$ values as
	indicated. The unfilled (filled) symbols represent the equilibrium (nonequilibrium) curves.
	The solid line represent the exact equilibrium curve and dotted lines are a guide to eyes. }

\end{figure}
}

\begin{document}
\title{Hysteresis and nonequilibrium work theorem for DNA unzipping}

\author{Rajeev Kapri}
\email{rkapri@iisermohali.ac.in}
\affiliation{Indian Institute of Science Education and Research Mohali,
Knowledge City, Sector 81, SAS Nagar -- 140 306, Punjab India.}

\begin{abstract}

	We study by using Monte Carlo simulations the hysteresis in unzipping and rezipping of a
	double stranded DNA (dsDNA) by pulling its strands in opposite directions in the fixed force
	ensemble. The force is increased, at a constant rate from an initial value $g_0$ to some
	maximum value $g_m$ that lies above the phase boundary and then decreased back again to
	$g_{0}$. We observed hysteresis during a complete cycle of unzipping and rezipping. We
	obtained probability distributions of work performed over a cycle of unzipping and rezipping
	for various pulling rates. The mean of the distribution is found to be close (the difference
	being within $10\%$, except for very fast pulling) to the area of the hysteresis loop. We
	extract the equilibrium force versus separation isotherm by using the work theorem on
	repeated non-equilibrium force measurements. Our method is capable of reproducing the
	equilibrium and the non-equilibrium force-separation isotherms for the spontaneous rezipping
	of dsDNA.

\end{abstract}

\pacs{87.14gk, 87.15.Zg, 36.20.Ey }

\date{\today}

\maketitle

\section {Introduction}

Unzipping of a double stranded DNA (dsDNA), an essential step in biological processes like DNA
replication and RNA transcription, is carried out by enzymes that exert an external force on
the strands of the DNA~\cite{Watson2003}. The phenomenon has been studied both
theoretically~\cite{BhattacharjeeJPA2000,LubenskyPRL2000} and
experimentally~\cite{BockelmannBiophysJ2002,DanilowiczPRL2004} by applying a pulling force on
the strands of the DNA.  In theoretical models, the strands of the DNA are modelled either on
the lattice, by random or self-avoiding walks, or in the continuum by worm like chains. It was
found that when a pulling force is applied on a dsDNA, the two strands unzip if the force
exceeds a critical value.  Below the critical force the DNA is in the zipped phase while above
it the DNA is in the unzipped phase. The force can be applied on the DNA by either keeping the
separation between the strands fixed (fixed distance ensemble) or by applying a fixed pulling
force (fixed force ensemble) on the strands. For the later case, the separation between the
strands fluctuates while it is the force needed to keep the separation fluctuates in the fixed
distance ensemble. The unzipping of dsDNA by a pulling force is a first-order phase transition
\cite{BhattacharjeeJPA2000,LubenskyPRL2000}. 

In a continuous phase transition large fluctuations in the order parameter are present near the
transition region that act as a precursor that something unusual is about to occur.  In the
case of DNA, if the melting is continuous, there will be large  fluctuations in the size and
shape of the denatured bubbles along the chain. These fluctuations are absent in a first-order
phase transition and the order parameter changes abruptly as the phase boundary is crossed.
However, there is usually a hysteresis associated with the first order transition which causes
the change to occur at a point that is slightly displaced from the phase boundary. This is
because at a first order phase boundary the two phases can coexist and be separated by
interfaces.  The energy of the interface acts as a barrier between two phases. Hysteresis is
often linked to the dynamics of interfaces. Some aspects of interfaces in DNA have been
discussed in Ref.~\cite{SadhukhanEPL2011}. Near the phase boundary, there is a region of
metastability where the system can stay in its previous phase even after crossing the phase
boundary. From the dynamics point of view, the relaxation time or the time scale to cross the
barrier becomes large near the transition and therefore, there is a conflict between relaxation
and the time scale of change of parameters. This produces hysteresis.  A classic example of
first-order transition in which hysteresis has been studied in detail is the Ising model below
the critical temperature in an external magnetic field~\cite{ChakrabartiRMP1999}. In recent
years, hysteresis has been studied in unbinding and rebinding of biomolecules under a pulling
force by using single molecule manipulation techniques
\cite{HatchPRE2007,FriddleJPCC2008,TshiprutJCP2009} because it can provide useful information
on kinetics of conformational transformations, potential energy landscape, and controlling the
folding pathway of a single molecule \cite{LiPNAS2007}.

The equilibrium statistical mechanics is a celebrated framework that gives the microscopic
description of the thermodynamics of the system. However, one major challenge, often faced in
designing experiments, is the requirement of thermodynamic equilibrium; the system should
remain in equilibrium, or at least at quasi-equilibrium, throughout the course of the
experiment, and needs to be equilibrated whenever system parameters are changed. However, in
the last decade many remarkable identities, known as nonequilibrium work or fluctuation theorems
(see Ref.~\cite{EvansAP2002} for a review), are developed that bridges the gap between the
nonequilibrium and equilibrium statistical mechanics. One of them is the Jarzynski
equality (JE)~\cite{JarzynskiPRL1997}, which connects the thermodynamic free-energy differences
between the two equilibrium states (say $A$ and $B$), $\Delta F = F_B - F_A$, and the
irreversible work done, $W$, in taking the system from one equilibrium state $A$ to a
non-equilibrium state having the same external conditions as that of the other equilibrium
state $B$. The relation between $\Delta F$ and $W$ is
\begin{equation}\label{Eq:JE}
	e^{- \Delta F / k_B T } = \langle e^{ - { W }/{k_B T} } \rangle,	
\end{equation}
where $k_B$ is the Boltzmann constant and $T$ is the absolute temperature. The bracket $\langle
\cdots \rangle$ denotes average over all possible paths between $A$ and $B$. Recently,
Sadhukhan and Bhattacharjee \cite{SadhukhanJPA2010} have given a dynamics-independent
nonequilibrium path integral formulation of the JE that can easily be
generalized to cases involving parameters like temperature and interactions. The
applicability of JE has been questioned for boundary-switching processes. It
was shown that for spontaneous irreversible processes (e.g., the adiabatic expansion of the gas
into vacuum) the JE is not satisfied~\cite{SungPRE2008}. However, our simulations show that one
can get the equilibrium information by using JE and the multiple histogram
technique~\cite{FerrenbergPRL1989} even for the spontaneous rezipping of dsDNA.

In this paper we study the hysteresis in unzipping and rezipping of a homopolymer dsDNA when
its strands are pulled in opposite directions by a force. The force $g$ is increased, at a
constant rate $ \dot{g} \equiv \Delta g / \Delta t$, from some initial value $g_{0}$ to some
maximum value $g_{m}$ that lies above the phase boundary. The force $g$ is then decreased back
to $g_{0}$ at the same rate. We observed hysteresis during a complete cycle of unzipping and
rezipping. By using the work theorem on repeated non-equilibrium force measurements on the
forward and the backward paths, we extract the equilibrium force-distance isotherm.  We also
show that our procedure can be used to obtain the complete equilibrium and nonequilibrium
force-separation isotherms for the spontaneous rezipping of dsDNA. In this process, the dsDNA
is initially at equilibrium with a maximum pulling force $g=g_m$ and the force is suddenly
decreased to a minimum force $g=g_0$.

The paper is organized as follows: In Sec.~\ref{sec:model}, we define our model and the details
of the Monte Carlo simulations. We compare the equilibrium results obtained by the simulation
with the exact results known for the model. The results are discussed in Sec.~\ref{sec:results}
and summarized in Sec.~\ref{sec:summary}.

\section{Model} \label {sec:model}

The two strands of a homo-polymer DNA are represented by two directed self-avoiding walks on a
$d=1+1$ dimensional square lattice.  The walks starting from the origin are restricted to go
towards the positive direction of the diagonal axis ($z$-direction) without crossing each
other. The directional nature of the walks takes care of self-avoidance and the correct base
pairing of DNA, that is, the monomers which are complementary to each other are allowed to occupy
the same lattice site. For each such overlap there is a gain of energy $-\epsilon$ ($\epsilon
>0$).  One end of the DNA is anchored at the origin and a force $g$ acts along the transverse
direction ($x$-direction) at the free end. The schematic diagram of our model is shown in
Fig.~\ref{fig:model}.

\figOne

Let $D_n(x)$ be the partition function in the fixed distance ensemble of a dsDNA of length $n$
with separation $x$ between the $n$th monomers of two strands. $D_n(x)$ satisfies the following
recursion relation:
\begin{eqnarray}	\label{eq:recrel}
	D_{n+1}(x) &=& \left[ D_n(x+1) + 2D_n(x) + D_n(x-1) \right]  \cr
						 & & \qquad \qquad \qquad \times \left[ 1 + \left( e^{\beta \epsilon} - 1\right) 
						 \delta_{x,0} \right],
\end{eqnarray}
with an initial condition $D_0(x)=e^{\beta \epsilon} \delta_{x, 0}$. 

The above recursion relation can be solved exactly via the generating function technique
\cite{MarenduzzoPRE2001,MarenduzzoPRL2002,KapriPRL2004}. The aim is to calculate the
singularities of the generating function. The singularity closest to the origin gives the phase
of the DNA and the phase transition takes place when two singularities cross each other. The
generating function for $D_n(x)$ can be taken of the form (ansatz)
\begin{equation}
	\hat{D}(z,x) = \sum_n z^n D_n(x) = \lambda^{x}(z) A(z).
\end{equation}
When used in the above recursion relation, we obtain $\lambda(z) = (1- 2z -\sqrt{1-4z})/(2z)$
and $A(z) = 1/ \left[ 1 - z \left( 2 + \lambda(z) \right) e^{\beta \epsilon}  \right]$. The
singularities of $\lambda(z)$ and $A(z)$ are $1/2$ and $z_2 = \sqrt{1 -e^{-\beta \epsilon} } -
1 + e^{-\beta \epsilon}$ respectively. The zero force melting, which comes from $z_1 = z_2$,
takes place at a temperature  $T_m = \epsilon/\ln (4/3)$. In the large length limit, $D_n(x)$
can be approximated as $D_N(x) \approx \lambda^x(z_2) / z_2^{N+1}$, with the free energy $\beta
F = N \ln z_2 - x \ln \lambda (z_2)$. The force required to maintain the separation $x$ is then
given by
\begin{equation}
	g(T) = \frac{\partial F}{2\partial x} = - \frac{k_B T}{2} \ln \lambda (z_2),
	\label{eq:gc}
\end{equation}
where $2$ in the denominator is the unit length of the diagonal of the square lattice. In the
fixed force ensemble, the generating function can be written as
\begin{eqnarray}
	\mathcal{G}(z, \beta, g) &=& \sum_x e^{2\beta g x} \sum_{n} z^n D_n(x) = \sum_x e^{2 \beta g x}
	\lambda^x(z) A(z) \cr
	&=& \frac{A(z)}{1 - \lambda(z) e^{2\beta g}},
\end{eqnarray}
which has an additional force dependent singularity $z_3 = 1/[2 + 2 \cosh (2\beta g)]$. The
phase boundary comes from $z_2=z_3$, and is given by
\begin{equation}\label{eq:gcff}
	g(T) = \frac{k_B T}{2} \cosh^{-1} \left [ \frac{1}{2} \frac{1}{ \sqrt{1 -e^{-\beta
	\epsilon} } - 1 + e^{-\beta \epsilon}} -1 \right ],
\end{equation}
which is same as the phase boundary obtained in the fixed distance ensemble. The phase diagram
for $k_B = 1$ is shown in Fig.~\ref{fig:pd}(a). The phase diagrams of DNA unzipping has been
obtained previously in Refs. \cite{MarenduzzoPRE2001,MarenduzzoPRL2002,KapriPRL2004}.

\figTwo

For many other equilibrium properties, based on thermal averaging, the exact transfer matrix
technique can be used to obtain numerically the partition function $D_N(x)$ for the DNA of
length $N$. This technique gives exact numerical results for finite $N$, which is rather
difficult in the generating function method. To obtain $D_N(x)$ at any given temperature we
start with an initial condition $D_0(0) = 1$ and iterate the above recursion relation [i.e.
Eq.(\ref{eq:recrel})] $N$ times. The equilibrium average separation between the end monomers,
$\langle x \rangle_{\rm eq}$, can then be obtained by 
\begin{equation}\label{eq:avx}
	\langle x \rangle_{\rm eq} = \frac{ \sum_x x \ D_N(x) e^{\beta g x}}{ \sum_x D_N(x)
	e^{\beta g x} },
\end{equation}
which is shown in Fig.~\ref{fig:pd}(b) for a DNA of length $N=128$ at $T=1.0$.

In this paper, to go beyond equilibrium, we perform Monte Carlo simulations of the model by
using the Metropolis algorithm, which allows us to study both equilibrium and off-equilibrium
behaviors. The strands of the DNA undergo Rouse dynamics that consists of local corner-flip or
end-flip moves~\cite{Doi1986} that do not violate mutual avoidance (the self-avoidance is taken
care by the directional nature of the walks). The elementary move consists of selecting a
random monomer from a strand, which itself is chosen at random, and flipping it. If the move
results in overlapping of two complementary monomers, thus forming a base-pair between the
strands, it is always accepted as a move.  The opposite move, that is, the unbinding of
monomers, is chosen with the Boltzmann probability $\eta = \exp(- \epsilon/ k_B T)$. If the
chosen monomer is unbind, which remains unbind after the move is performed is always accepted.
The time is measured in units of Monte Carlo steps (MCSs).  One MCS consists of $2N$ flip
attempts, that is, on an average, every monomer is given a chance to flip. Throughout the
simulation, the detailed balance is always satisfied. From any starting configuration, it is
possible to reach any other configuration by using the above moves.  Throughout the paper,
without loss of generality, we have chosen $\epsilon = 1$ and $k_B = 1$.

To check if the results obtained by using the above mentioned moves are consistent with the
analytical results obtained previously, we calculate the force $g$ vs equilibrium average
separation $\langle x \rangle_{\rm eq}$ between the end monomers of the DNA of length $N=128$ at
$T=1$. This is shown in Fig.~\ref{fig:pd}(b) by filled circles. Every data point is obtained by
first equilibrating the system for $2 \times 10^5$ MCSs and then averaged over $10^4$ different
realizations. In the same plot we have also shown, by solid line, the force-distance isotherm
obtained by using the exact transfer matrix calculations for the model. The results of Monte
Carlo simulations match excellently with the exact result. The equilibrium configurations of
the dsDNA of length $N=128$ at temperature $T=1$ for two different forces $g=0.65$, which lies
just below the phase boundary, and $g=0.9$, which is far away from the phase boundary (and also
the maximum force used in this paper at $T=1$) are also shown in Fig.~\ref{fig:pd}(c).  These
configurations show that the DNA is in the zipped phase (with a small Y-fork at the end) for
the force below the critical force and in the unzipped phase for the force above the critical
force.

To study the hysteresis in DNA, we start the simulation with a valid configuration of a dsDNA
of length $N=128$ at $T=1$ and $N=256$ at $T=3.6$. The later temperature is above the melting
temperature $T_m \approx 3.476$ of the dsDNA for the model used in this paper. The system is
first equilibrated with zero pulling force $g_{0} =0$. The force $g$ is incrementally increased
from $g_0$ to $g_m = 0.9$ at $T=1$ ($g_m = 1.0$ is used at $T=3.6$) at a step of $\Delta g^{F}
= 0.01$ by using the following protocol
\begin{equation}
	g_{i}^{F} = g_0 + i \Delta g^{F},
\end{equation}
where $i = 0, 1, 2, \ldots, n$ with $n = (g_m - g_0)/\Delta g^{F}$ is the number of steps
between the initial and the final force values. The superscript $F$ denotes
the forward path. For the backward path (denoted by superscript $B$) the force is
incrementally decreased from $g_m$ to $g_0$ by $\Delta g^{B} = - \Delta g^{F}$. The number of
steps $n$ and the time interval $\Delta t$ are kept same as that of the forward path.

Each step of the process can be thought of two substeps. In the first substep, the force is
increased by $\Delta g^{F}$. Therefore, an amount of work $\Delta W^F = - \Delta g^{F} x_{i}^F$
has to be performed on the system, where $x_{i}^F$ is the separation between the end monomers of
the DNA at the beginning of the $i$th step. In the second step, the system is relaxed for the
time interval $\Delta t$ in the presence of the pulling force $g_{i+1}$. The total work
performed on the system during the complete forward process is
\begin{equation}
	\label{eq:workF}
	W^{F} = - \Delta g^{F} \sum_{i=0}^{n-1} x_{i}^F.
\end{equation}
Similarly, for the backward path the work performed by the system is
\begin{equation}
	W^{B} = - \Delta g^{B} \sum_{i=0}^{n-1} x_{i}^B.
	\label{eq:workB}
\end{equation}
The above procedure is repeated many times to obtain various trajectories. For each
realization, the system is initially equilibrated at $g_0$ but no attempt has been made to
equilibrate the system at the maximum force $g_m$ used in this paper. The total work done over
a complete unzipping and rezipping cycle is given by the sum of the work performed along the
forward and the backward paths
\begin{equation}
	W = W^F + W^B.
\end{equation}
The work performed $W$ is different for different realizations. The following sign convention
is adopted in this paper: The positive (negative) sign of work denotes the work done on the
system (by the system).

\section{Results and Discussions} \label{sec:results}
 
Before discussing our results let us fix some notations to avoid confusion. We are working in
the fixed force ensemble and define the pulling rate by $\dot{g} \equiv \Delta g / \Delta t$.
We fix the force interval to $\Delta g = 0.01$ (in magnitude) for both the forward and the
backward paths and change the time interval $\Delta t$ to change the pulling rate. Therefore,
instead of giving the actual numerical value we just give the time interval $\Delta t$ for the
pulling rate. For example, the pulling rate for $\Delta t = 100$ means $\dot{g} \equiv 0.01/100
= 10^{-4}$.

\subsection{Hysteresis curves}

\figThree

In Fig.~\ref{fig:sam}(a), we have shown, for two different realizations, the force $g$ versus
separation $x$ between the end monomers of dsDNA for the pulling rate for $\Delta t = 1000$.
The forward and backward paths are shown respectively by open and filled symbols. These
realizations reveal that the system does not get enough time to relax to the equilibrium and
shows hysteresis. The average separation $\langle x \rangle$ at force $g$ is obtained by
\begin{equation} 
	\langle x \rangle = \frac{1}{M} \sum_{i=1}^{M} x_{i},
\end{equation} 
for both the forward and backward paths. The resulting hysteresis for various pulling rates
averaged over $M=10^5$ realizations are shown in Fig.~\ref{fig:sam}(b). When $\Delta t$ is
smaller (i.e., the pulling rate is higher), the system does not get enough time to respond to
the change in the pulling force and the average separation between the strands is small even if
the pulling force is greater than the critical force needed to unzip the DNA. On decreasing the
force from the maximum, the separation between the strands initially increases because the
force is still greater than the critical unzipping force and the system gets some more time to
relax. Therefore, on the backward path there exist a force at which the average separation is
equal to the equilibrium separation. This is the point at which the equilibrium curve cuts the
backward path of the hysteresis loop in Fig.~\ref{fig:sam}(b). On decreasing the force further,
the average separation decreases slowly and the system is again driven away from the
equilibrium. If we join the forward and the backward paths we get a small hysteresis loop. The
area of the loop gives the amount of heat that is deposited to the system. Since
for the fast pulling rate (i.e., small $\Delta t$) only a small segment of the dsDNA can be
opened by the force, the area of the loop is small. As $\Delta t$
is increased, the system gets more and more time to respond to the change in the pulling force and the average separation between the strands at $g_m$ increases.
This increase in separation continues initially on the backward path also and therefore the
area of the hysteresis loop increases as is visually seen from Fig.~\ref{fig:sam}(b). But
the area of the loop cannot increase forever with the increase of $\Delta t$. For sufficient
large $\Delta t$, the average separation between the strands at $g_m$ becomes closer to the
equilibrium separation as can be seen for $\Delta t = 10000$ curve. We will see later that
for such cases, the nonequilibrium measurements along the backward paths can also be used to
calculate the equilibrium curve. If $\Delta t$ is very large, that is, the pulling is very slow,
the system gets sufficient time to get equilibrated before the force is increased to a new
value.  Therefore, the system remains in equilibrium for all force values and does not show any
hysteresis (zero loop area). This situation is shown in Fig.~\ref{fig:sam}(b) by filled
circles. The area of the hysteresis loop calculated by integrating numerically using
trapezoidal rule for various $\Delta t$ values are tabulated in Table I, which confirms the
above statement. 

\begin{table}[h]
\begin{center}
	\begin{tabular}{ l  c  c  c }
		\hline
		\hline
		$\Delta t$ & $\langle W \rangle$ & $\sigma$ & Area of \\
		           &                     &           &hysteresis loop \\
		\hline

		100 	& $10.3 \pm 0.3$    & $8.0 \pm 0.3$ & 12.636 \\
		500 	& $25.6 \pm 0.2$    & $12.8 \pm 0.2$ & 28.057 \\
		1000 	& $37.5 \pm 0.2$    & $14.3 \pm 0.2$ & 38.066 \\
		5000 	& $26.17 \pm 0.03$  & $6.75 \pm 0.03$ & 27.687 \\
		10000 & $16.33 \pm 0.02$  & $5.47 \pm 0.02$ & 18.029 \\  
		12000 & $14.42 \pm 0.05$	& $5.12 \pm 0.05$ & 16.138 \\
		\hline
		\hline

	\end{tabular}
\end{center}
	
\caption{ The average $\langle W \rangle$ and the standard deviation $\sigma$ of the
probability distribution of work performed over an unzipping and rezipping cycle, and the area
of the hysteresis loop for various $\Delta t$ values.  }

\end{table}

\subsection{Probability distribution of work performed over a cycle}

As stated in Sec.~\ref{sec:model}, the work performed during a complete unzipping and rezipping
cycle are different for different realizations. There are a few realizations for which the work
obtained during the rezipping process is more than the work performed during the unzipping
process. One such realization is shown in Fig.~\ref{fig:sam}(a) by circles. The second law of
thermodynamics cannot be violated. But if one looks at an individual or a group of such
trajectories, there is an illusion of a violation. However, for the majority of trajectories,
which look more or less like the realization shown in Fig.~\ref{fig:sam}(b) by diamonds, the
work performed during the unzipping process is more than the work obtained during the rezipping
process.  The probability distribution of work guarantees that the second law of thermodynamics
is satisfied on an average.

\figFour

In Fig.~\ref{fig:prob}, we have plotted the probability distribution of work, $P(W)$, performed 
over a complete unzipping and rezipping cycle for various $\Delta t$ values. The solid lines
are the Gaussian fit to the data
\begin{equation}
	P(W) = A \exp \left[ - \frac{ (W - \langle W \rangle )^2}{ 2
	\sigma^2} \right], 
\end{equation}
where, $\langle W \rangle$ and $\sigma$, respectively, represent the average work performed
during a cycle and the standard deviation of the distribution and $A$ is the normalization
constant. The values obtained for various $\Delta t$ are tabulated in Table I.

One can observe from the figure that for $\Delta t = 100$ (i.e., for faster pulling rates), the
probability distribution deviates from the Gaussian distribution. The asymmetry of the
distribution is quite visible, so we have not shown the Gaussian fit for this data but joined
data points by a dashed line to guide the eyes. The distribution is peaked towards lower $W$ values
and therefore the probability of obtaining negative work over a cycle is also higher. The
average work is however positive. As the pulling rate decreases, the distribution becomes more
and more symmetric and tend towards the Gaussian distribution. The peak of the distribution
first shifts towards higher $W$ values and it becomes broader as can be seen in the figure for
$\Delta t = 1000$. On decreasing the pulling rate further, the mean and the width of the
distribution again starts decreasing.  If the pulling rate is extremely slow, the system
remains in equilibrium at all times during the forward and the backward path, and therefore the
work done on the system during the forward path is exactly equal to the work done by the system
during the backward path. The total work performed over a cycle is zero and the
distribution $P(W)$ is sharply peaked at $W=0$. Therefore, at equilibrium, the average work done
on the system is exactly equal to the area of the hysteresis loop. In the nonequilibrium
regime, except for very fast pulling rate (i.e., $\Delta t = 100$), the difference of the
average work done on the system and the area of the hysteresis loop is within $10\%$ (see Table
I) of the area of the loop. The difference comes because only $10^5$ trajectories
are simulated. This many trajectories are good enough to calculate the average extensions and
hence the area of the hysteresis curves. However, these are not good enough to calculate the
probability distributions of the work, which are obtained by binning the data. If more
realizations are used in obtaining the probability distributions the difference between the two
quantities will become smaller. 

For a cyclic process, the initial and the final equilibrium states are the
same. Substituting $\Delta F = 0$ in Eq.~(\ref{Eq:JE}) one gets
\begin{equation}\label{eq:JEC}
 Z \equiv	\langle e^{ - \beta W } \rangle \equiv \frac{1}{M} \sum_{i=1}^{M} e^{-\beta W_i} = 1.
\end{equation}
This equality strictly holds for infinite trajectories $M \to \infty$. The convergence is
often slow because it picks up exponential weights and the trajectories that contribute to the
sum (i.e. having extreme weights) are rare. We calculate $Z$ for $\Delta t = 10000$ case. On
averaging over $M = 10^2, 10^3, 10^4$, and $10^5$ trajectories, we obtain $Z = 0.0027, 0.0102,
0.0827$, and $0.2784$, respectively. This shows that $Z$ increases with number of trajectories
used in averaging. This direct sum over paths may not be a useful numerical way to estimate the
free-energy or establish Eq.~(\ref{eq:JEC}). However, a weighted average as discussed below is
more efficient for averages. The probability distributions $P(W)$ obtained above can be used to
calculate $Z$:
\begin{equation}\label{eq:WA}
	Z = \int_{-\infty}^{\infty} dW e^{-\beta W} P(W) = \exp \left( - \beta \langle W \rangle +
	\frac{\beta^2 \sigma^2}{2} \right).
\end{equation}
Substituting the value of $\langle W \rangle$ and $\sigma$ for $\Delta t = 10000$ from
Table I, we get $Z = 0.2542$, which is somewhat smaller than the value obtained from
Eq.~(\ref{eq:JEC}) for $10^5$ trajectories. Equation~(\ref{eq:WA}) is not applicable for fast
pulling (i.e., smaller $\Delta t$ values) because of deviations from Gaussian. For those cases,
higher cumulants are important. In the limit of infinite trajectories, $P(W)$ will
become the true probability distribution of work and the value of $Z$ obtained by either
methods will approach $1$.

\subsection{Equilibrium curves from non-equilibrium measurements}

In this section, we discuss the procedure that can be used to obtain the equilibrium
force-distance isotherms by using non-equilibrium measurements on the forward and the backward
paths. Our technique is similar to that of Hummer and Szabo~\cite{HummerPNAS2001}, but in the
conjugate ensemble). The above method (Ref.~\cite{HummerPNAS2001}) has been used successfully
to obtain the zero force free energy on single molecule pulling experiments in constant
velocity ensemble~\cite{GuptaNPhys2011}. 

\figFive

In Figs.~\ref{fig:hyst}(a), \ref{fig:hyst}(b) and \ref{fig:hyst}(c), we have plotted the
hysteresis loops for $\Delta t = 5000$, $10000$, and $12000$, and the equilibrium curve
obtained by using the exact transfer matrix for $N=128$ and $T=1$, which is below the melting
temperature $T_m = 3.476$.  In the same plot we have also shown the equilibrium data calculated
from the nonequilibrium force measurements by using the procedure described below.

We divide the forward path into intervals of sizes $\Delta g$. Let $i$ and $k$ represent
respectively the indices for the sample and the force. The irreversible work done over the
$i$th non-equilibrium path, taking $W_{i0} = 0$ at $g_0$, is given by
\begin{equation}
	W_{ik} = - \Delta g \sum_{j=0}^{k} x_{ij}.
\end{equation}
By using $\exp(-\beta W_{ik})$ as the weight for path $i$, the equilibrium separation between
the end monomers of the dsDNA, $x_{k}^{\rm eq}$, at force $g_k$ can be obtained by
\begin{equation}
	\label{Eq:eqlm}
	x_{k}^{\rm eq} =  \frac{  \sum_{i=1}^{M} x_{ik} \exp \left( - \beta W_{ik}
	\right) } { \sum_{i=1}^{M} \exp \left( - \beta W_{ik} \right) }.
\end{equation}
The above procedure has been used by Sadhukhan and Bhattacharjee \cite{SadhukhanJPA2010} to
obtain the equilibrium curve. 

\figSix

In our simulation, we have $10^5$ $x_{k}$ values, at each force $g_k$. These values can be
used in Eq.~(\ref{Eq:eqlm}) to obtain $x_{k}^{\rm eq}$ at $g_k$. However, one can do better than
this. For a given temperature $T$ and force $g$, the system at equilibrium samples a narrow
phase space given by the Boltzmann distribution. The distributions at nearby force values
overlap with each other. The overlapping distributions can be properly weighted to obtain the
approximate density of states (DOS) of the system. This goes by the name of multiple histogram
technique \cite{FerrenbergPRL1989} and has been exploited in simulations. Once the DOS is
known, the observables can be calculated at any other force. To achieve this we build up the
histogram $H_k(x)$ at force $g_k$. For the $i$th realization, if the separation is $x$, we
increment the corresponding histogram value by $\exp(-\beta W_{ik})$,
\begin{equation}
	H_k(x) = \sum_{i=1}^{M} e^{-\beta W_{ik}} \ \delta_{x, x_{i}},
\end{equation}
where, $\delta_{x,x_{i}} = 1$  if $x=x_{i}$, and zero otherwise. The partition function $Z_k$ at
force $g_k$, obtained by using the multiple histogram technique~\cite{FerrenbergPRL1989}, reads
as
\begin{equation}
	\label{Eq:partfn}
	Z_k = \sum_{x} \rho(x) \exp(\beta g_k x), 
\end{equation}
where
\begin{equation}
	\rho(x) = \frac{ \sum_j \frac{H_j(x)}{Z_j} }{ \sum_j \frac{\exp(\beta g_j x)
	}{Z_j}},
	\label{Eq:dos}
\end{equation}
is the DOS. Equation (\ref{Eq:partfn}) needs to be evaluated self-consistently. We take initial
$Z_k$'s as
\begin{equation}
	Z_k =  \frac{1}{M} \sum_{i=1}^{M} \exp \left( - \beta  W_{ik} \right),
\end{equation}
and iterate Eq.~(\ref{Eq:partfn}) till it converges to a true DOS. The initial values are
motivated by the JE [see Eq.~(\ref{Eq:JE})]. By using $\rho(x)$, we can then
evaluate the equilibrium separation $x_{k}^{\rm eq}$ at force $g_k$ by
\begin{equation}
	x_{k}^{\rm eq} = \frac{1}{Z_k} \sum_{x} x \rho(x) \exp( \beta g_k x). 
	\label{Eq:xeq}
\end{equation}
The equilibrium force-separation curve obtained by using the data of $10^5$ nonequilibrium
forward paths for $\Delta t = 5000$ and $10000$, and $10^4$ paths for $\Delta t = 12000$, in
above procedure is shown by open symbols in Figs.~\ref{fig:hyst}(a)- \ref{fig:hyst}(c). The data for $\Delta t = 10000$ and $12000$ match reasonably well with the equilibrium curve obtained by the exact
transfer matrix method. The same procedure can also be adopted for the backward path.  The
equilibrium force-extension curve obtained by using the nonequilibrium data for the backward
path is shown by filled circles which match excellently with the exact curve for $\Delta t =
10000$ and $12000$. For the fast pulling case, that is, $\Delta t = 5000$, the results from the
forward paths match all other points except at the transition region due to poor statistics in
that region. The multihistogram approach, by analyzing the errors in the estimates of the
partition function, tries to correct for the unexplored part of the phase space.  Even for this
to work, one needs good sampling or better statistics. The other approach is to generate rare
conformations that have dominant contributions in the weighted sum by using special
algorithms~\cite{SunJCP2003,YtrebergJCP2004,OberhoferPRE2007}. It would be interesting to do a
comparative study of the two approaches. Special algorithms are important in
generating trajectories for very fast pulling rates (e.g., $\Delta t= 100$, $500$, etc.), where
not enough rare configurations can be generated by simply including more trajectories due to
low probabilities of such configurations. We have also shown, in Figs.~\ref{fig:hyst}(d)- \ref{fig:hyst}(f),
the hysteresis and equilibrium curves obtained by using the above procedure for $N=256$ at
temperature $T=3.6$ which is above the melting temperature of the DNA.  These are obtained for
the pulling rates with $\Delta t = 1000$, $2000$ and $3000$ respectively by averaging over
$10^4$ nonequilibrium paths. The equilibrium curve obtained by using the data of the forward
paths match excellently with the exact curve while the curve that uses the data of the backward
paths deviates from the exact curve at higher forces. As stated previously, we have not made
any attempt to equilibrate the system at the maximum force $g_m$. Still, we could get an
excellent match with the equilibrium curve by using the data for the backward paths for $T=1$
but not with the data for $T=3.6$.  This can be understood by observing the hysteresis curves
near $g_m$ for both cases. For $T=1$, with $\Delta t = 10000$ and $12000$, the average
separation between the end monomers of the DNA at $g_m$ is quite closer to the equilibrium
curve.  Therefore, the system is practically in equilibrium at the beginning of the backward
paths and one can use the work theorem [Eq.~(\ref{Eq:JE})] to obtain equilibrium properties.
However, for $T=3.6$, the system has not reached the equilibrium at $g_m$ and so the
requirement of work theorem, that is, the system should initially be in equilibrium, is not
satisfied for the backward paths and it cannot be applied for this case. If we allow initial
equilibration at the maximum force $g_m$ then the work theorem can be applied for the backward
paths also and the weighted data on the backward paths should give us the equilibrium curve.
This we have explored at $T=3.6$ for various $\Delta t$ values in
Fig.~\ref{fig:hysteq}. One can see from the figure that for $\Delta t = 500$, $1000$, and
$2000$, the curves obtained by using our procedure on $10^4$ nonequilibrium trajectories match
reasonably well with the exact curve.  For $\Delta t = 100$, which is the fastest pulling rate
reported in this paper, the curves deviate from the exact curve at other ends but this is due
to the poor statistics in that region. In Fig.~\ref{fig:hysteq100}, we have shown how these
curves behave when the weighted average is taken over a different number ($10^3$, $10^4$ and
$10^5$) of trajectories. On including more trajectories, we can see that the curves move
towards the exact equilibrium curve. To get better statistics in the fast pulling regime, one
can use algorithms~\cite{SunJCP2003,YtrebergJCP2004,OberhoferPRE2007} mentioned previously to
generate rare trajectories. These can be included in the above procedure to get a better match
but we have not attempted it in this paper. It is clear from the above discussion that the
above procedure can be used successfully to obtained the equilibrium force-separation isotherms
from the nonequilibrium measurements when a pulling force is applied on the strands of the
dsDNA that varies with a constant rate.  

\figSeven

\figEight 

To check the robustness of the above method, we study the spontaneous rezipping of dsDNA. The
DNA is initially equilibrated with a pulling force $g_m = 0.725$ at a temperature $T=1$. The
force $g_m$ lies above the phase boundary (but close to it) and the dsDNA is in the unzipping
phase.  The force is suddenly decreased to a lower value $g_0$ that lies below the phase
boundary in which the DNA is in the zipped phase (i.e., a different boundary condition than
$g_m$). From the initial to the final force, the system is equilibrated for $1$ MCS (i.e.,
$\Delta t = 1$).  It is interesting to know if the method discussed above is capable of
predicting the equilibrium and the nonequilibrium force-separation curves at various other
force values. In Fig.~\ref{fig:spont} we have shown the curves obtained by using the above
method with $10^4$ trajectories when the force is spontaneously decreased from $g_m = 0.725$ to
$g_0=0.6$, $0.4$, $0.2$, and $0.0$. The equilibrium (nonequilibrium) curves are shown by the
unfilled (filled) symbols. An excellent match with the exact force-separation isotherm shows
that the above method can be used to predict the equilibrium force-separation isotherms from
the spontaneous rezipping of dsDNA.  The application range of the above method is limited by
how good statistics of the end separation one can get. The maximum force $g_m = 0.725$ is
chosen near the phase boundary because, even though the DNA is in the unzipped phase, many
configurations exist, which is needed to sample the whole phase space. If we take $g_m$ far
away from the phase boundary, the unzipped phase will only have fully stretched configurations
and isotherms extracted by the above method will deviate from the exact result at smaller force
values due to poor statistics in that region. The fact that the spontaneous unzipping is
capable of reproducing the exact equilibrium force-separation isotherms makes the above method
an excellent candidate for analyzing the pulling experiments. It, however, needs to be explored
further. The detailed analysis will be published elsewhere.  

\section{Concluding Remarks} \label{sec:summary}

To summarize, we have studied the hysteresis in unzipping and rezipping of a dsDNA in the fixed
force ensemble. We found that the area of the hysteresis loop depends on the pulling rate. For
fast pulling the area of the loop is smaller. On decreasing the pulling rate, the area of the
loop first increases, and then starts decreasing due to the system's proximity to the
equilibrium for the sufficiently slow pulling rates. On decreasing the pulling rate further,
the system remains in equilibrium at all intermediate force values and the area of the loop
becomes zero. We obtained the probability distributions of work performed over a complete
unzipping and rezipping cycle for various pulling rate. The average of this distribution is
found to be very close (the difference being within $10\%$, except for very fast pulling i.e.
$\Delta t = 100$) to the area of the hysteresis loop.  We also discussed a procedure to obtain
equilibrium force-distance isotherms by using repeated non-equilibrium measurements on the
forward paths. We found that if the pulling rate is such that the average separation between
the end monomers at the maximum force used is close to the equilibrium curve, the backward path
gives better results than the forward paths. Furthermore, using our technique,
one can obtain the complete equilibrium and the non-equilibrium force-separation isotherms for
the spontaneous rezipping of dsDNA. We believe that our multiple histogram based algorithm using
work theorem can be implemented in molecular manipulation machines to provide both the equilibrium
and the nonequilibrium information for pulling experiments.

\section*{Acknowledgements}

I thank Professor S. M. Bhattacharjee for comments and discussions. This work is supported by
DST Grant (Grant No.  SR/FTP/PS-094/2010).


\end{document}